\def\~{{$\tilde{\phantom{a}}$}}

\documentclass [12pt] {article}
\usepackage{epsfig}
\usepackage{color}

\def\thebibliography#1{\section{References}\markboth
 {REFERENCES}{REFERENCES}\list
 {[\arabic{enumi}]}{\settowidth\labelwidth{[#1]}\leftmargin\labelwidth
 \advance\leftmargin\labelsep
 \usecounter{enumi}}
 \def\newblock{\hskip .11em plus .33em minus -.07em}
 \sloppy
 \sfcode`\.=1000\relax}
\def\upcite#1{\raise6pt\hbox{\scriptsize
\cite{#1}}}
  \def\lsim{\mathrel {\vcenter {\baselineskip 0pt \kern 0pt
    \hbox{$<$} \kern 0pt \hbox{$\sim$} }}}
    \def\gsim{\mathrel {\vcenter {\baselineskip 0pt \kern 0pt
    \hbox{$>$} \kern 0pt \hbox{$\sim$} }}}

\def\hline{\noalign{\hrule \vskip2pt}}

\def\|{\ifmmode\Vert\else \char`\|\fi}
\ifx\oldzeta\undefined                          
  \let\oldzeta=\zeta                            
  \def\zzeta{{\raise 2pt\hbox{$\oldzeta$}}}     
  \let\zeta=\zzeta                              
\fi

\ifx\oldchi\undefined                           
  \let\oldchi=\chi                              
  \def\cchi{{\raise 2pt\hbox{$\oldchi$}}}       
  \let\chi=\cchi                                
\fi

\def\frac#1#2{{#1 \over #2}}

\def\half{\ifinner {\scriptstyle {1 \over 2}}
   \else {1 \over 2} \fi}

\def\simge{\mathrel{%
   \rlap{\raise 0.511ex \hbox{$>$}}{\lower 0.511ex \hbox{$\sim$}}}}
\def\simle{\mathrel{
   \rlap{\raise 0.511ex \hbox{$<$}}{\lower 0.511ex \hbox{$\sim$}}}}

\def\buildchar#1#2#3{{\null\!                   
   \mathop#1\limits^{#2}_{#3}                   
   \!\null}}                                    
\def\overcirc#1{\buildchar{#1}{\circ}{}}

\def\slashchar#1{\setbox0=\hbox{$#1$}           
   \dimen0=\wd0                                 
   \setbox1=\hbox{/} \dimen1=\wd1               
   \ifdim\dimen0>\dimen1                        
      \rlap{\hbox to \dimen0{\hfil/\hfil}}      
      #1                                        
   \else                                        
      \rlap{\hbox to \dimen1{\hfil$#1$\hfil}}   
      /                                         
   \fi}                                         %

\def\subrightarrow#1{
  \setbox0=\hbox{
    $\displaystyle\mathop{}
    \limits_{#1}$}
  \dimen0=\wd0
  \advance \dimen0 by .5em
  \mathrel{
    \mathop{\hbox to \dimen0{\rightarrowfill}}
       \limits_{#1}}}                           

\def\overlay#1#2{\ifmmode%
\setbox0=\hbox{$#1$}%
\setbox1=\hbox to\wd0{\hss$#2$\hss}\else%
\setbox0=\hbox{#1}%
\setbox1=\hbox to\wd0{\hss#2\hss}\fi%
#1\hskip-\wd0\box1 }

\def\pmb#1{\leavevmode\setbox0=\hbox{#1}%
\kern-.02em\copy0\kern-\wd0
\kern.04em\copy0\kern-\wd0
\kern-.02em\raise.04em\box0 }

\def\vereq#1#2{\lower3pt\vbox{\baselineskip1.5pt \lineskip1.5pt
\ialign{$\m@th#1\hfill##\hfil$\crcr#2\crcr\sim\crcr}}}

\def\tensor#1{\protect\@ontopof{#1}{\leftrightarrow}{1.15}\mathord{\box2}}
\def\overstar#1{\protect\@ontopof{#1}{\ast}{1.15}\mathord{\box2}}
\def\overdots#1{\protect\@ontopof{#1}{\cdots}{1.0}\mathord{\box2}}
\def\overcirc#1{\protect\@ontopof{#1}{\circ}{1.2}\mathord{\box2}}
\def\loarrow#1{\protect\@ontopof{#1}{\leftarrow}{1.15}\mathord{\box2}}
\def\roarrow#1{\protect\@ontopof{#1}{\rightarrow}{1.15}\mathord{\box2}}

\def\@ontopof#1#2#3{%
{\mathchoice
{\@@ontopof{#1}{#2}{#3}\displaystyle\scriptstyle}%
{\@@ontopof{#1}{#2}{#3}\textstyle\scriptstyle}%
{\@@ontopof{#1}{#2}{#3}\scriptstyle\scriptscriptstyle}%
{\@@ontopof{#1}{#2}{#3}\scriptscriptstyle\scriptscriptstyle}%
}%
}

\def\@@ontopof#1#2#3#4#5{%
\setbox0=\hbox{$#4#1$}%
\setbox1=\hbox{$#5#2$}%
\setbox2=\hbox{}\ht2=\ht0 \dp2=\dp0 %
\ifdim\wd0>\wd1 %
\setbox1=\hbox to\wd0{\hss\box1\hss}%
\mathord{\rlap{\raise#3\ht0\box1}\box0}%
\else   %
\setbox1=\hbox to.9\wd1{\hss\box1\hss}%
\setbox0=\hbox to\wd1{\hss$#4\relax#1$\hss}%
\mathord{\rlap{\copy0}\raise#3\ht0\box1}%
\fi
}%

\def\lambdabar{\protect\@lambdabar}
\def\@lambdabar{%
\relax
\bgroup
\def\@tempa{\hbox{\raise.73\ht0
\hbox to0pt{\kern.25\wd0\vrule width.5\wd0
height.1pt depth.1pt\hss}\box0}}%
\mathchoice{\setbox0\hbox{$\displaystyle\lambda$}\@tempa}%
{\setbox0\hbox{$\textstyle\lambda$}\@tempa}%
{\setbox0\hbox{$\scriptstyle\lambda$}\@tempa}%
{\setbox0\hbox{$\scriptscriptstyle\lambda$}\@tempa}%
\egroup
}

\def\corresponds{{\lower.2ex\hbox{=}}{\rm\kern-.75em^\triangle}}
\def\succsim{\succ\kern-.9em_\sim\kern.3em}
\def\precsim{\prec\kern-1em_\sim\kern.3em}
\def\slantfrac#1#2{\kern1em^{#1}\kern-.3em/\kern-.1em_{#2}}

\begin{document}

\begin{center}
{\Large\bf The Charge Distribution on the Cathode of a Straw Tube Chamber}
\\

\medskip

Changguo Lu and Kirk T.~McDonald
\\
{\sl Joseph Henry Laboratories, Princeton University, Princeton, NJ 08544}
\\
(Oct.~1, 1998)
\end{center}

\section{Problem}

A straw tube chamber is a low-cost version of a proportional counter.  These
devices consist of a pair of coaxial conducting cylinders with the region 
between the cylinders filled
with a gas such as argon.   The inner cylinder of
radius $a$ is the anode, and is held at potential $V$;
the outer cylinder of radius $b$ is the cathode, and is grounded.

If a penetrating charged particle passes through
the chamber, it will ionize about two gas molecules per mm of path length.
The ionization electrons are pulled by the electric field towards the anode.
Close to the anode, the field is strong enough that the electrons
gain enough energy during one mean free path to ionize the molecule they
hit next, liberating one or more additional electrons.  In a proportional
chamber, the field is kept low enough that the resulting Townsend avalanche
involves $10^4$-$10^6$ molecules.

What is the time dependence, $I(t)$, of the current that flows off the anode due
to the avalanche of a single initial electron?

What is the spatial dependence, $q(z)$ of the charge distribution induced on 
the anode during the time when the current is large, where the $z$ axis is the
chamber axis?   You may restrict your attention to values of $z$ far from the
ends of the tube of length $l$.

Measurement of the
charge distribution via a segmented cathode permits localization  in $z$ of
the ionization, and hence, of the initiating charged particle \cite{straw}.

You may ignore the tiny current that flows while the electron drifts towards the
anode.  The avalanche takes place so close to the anode, that the small
remaining drift time for the electrons to reach the anode may also be ignored.
In this approximation, the situation at $t = 0$ is that electrons of total 
charge $-q_0$ reside on
the anode in close proximity to positive ions of total charge $+q_0$.  Current
flows off the anode only when some of the field lines from the positive ions
detach from the electrons on the anode, and extend to the cathode where
charge is induced to terminate these field lines.  This occurs only as the
positive ions move away from the anode, with velocity related by
\begin{equation}
v = \mu E,
\label{eq1}
\end{equation}
where $\mu$ is the positive ion mobility.

\subsection{$I(t)$ via Reciprocity and Weighting Fields}

This problem can be solved by an application of
Green's reciprocation theorem, which states that if a set of fixed conductors 
is at potentials $V_i$ when carrying charges $Q_i$, and at potentials
$V'_i$ when carrying charges $Q'_i$, then
\begin{equation}
\sum_i V_i Q'_i = \sum_i V'_i Q_i.
\label{eq1a}
\end{equation}
To see this, we label the 3-dimensional potential distribution associated with
charges $Q_i$ by $\phi({\bf r})$, and that associated with charges $Q'_i$ by 
$\phi'$.  The space outside the conductors is charge free and with dielectric
constant $\epsilon = 1$.  Then $\nabla^2 \phi = 0 = \nabla^2 \phi'$ outside
the conductors.

We invoke Green's theorem (sec.~2.12 of \cite{Smythe}),
\begin{equation}
\int ( \phi \nabla^2 \phi' - \phi' \nabla^2 \phi) d{\rm vol}
= \oint (\phi \nabla \phi' - \phi' \nabla \phi) \cdot d{\bf S},
\label{eq1b}
\end{equation}
where we take the bounding surface $S$ to be that of the set of conductors.  
Hence,
\begin{equation}
0 = \sum_i \oint (V_i \nabla \phi'_i - V'_i \nabla \phi_i) \cdot d{\bf S}_i
= - 4 \pi \sum_i (V_i Q'_i - V'_i Q_i),
\label{eq1c}
\end{equation}
using Gauss' Law (in Gaussian units) that
\begin{equation}
4 \pi Q_i = \oint {\bf E_i} \cdot d{\bf S}_i 
= - \oint \nabla \phi_i \cdot d{\bf S}_i.
\label{eq1d}
\end{equation}

In the present problem, we have a small charge $q_0$ at position ${\bf r}_0(t)$
that moves under the influence of the field due to conductors $i = 1,...,n$
that are held at potentials $V_i$.  The charges $Q_i$ on the conductors obey
$Q_i \gg q_0$, so the motion of charge $q_0$ is determined, to a very good
approximation by the charges $Q_i$ on the conductors when $q_0 = 0$.  Hence,
the problem can be considered as the superposition of two situations:

A: charge $q_0$ absent; conductors $i = 1,...n$ at potentials $V_i$.

B: charge $q_0$ present; conductors $i = 1,...n$ grounded, with charges 
$\Delta Q_i$ on them.
\hfill\break\noindent
We are particularly interested in the charge on electrode 1, whose time rate of
change is the desired current $I(t)$. 

To use the reciprocation theorem, we suppose that in case B the charge resides
on a tiny conductor at position ${\bf r}_0$ that is at the potential
$V_0 = \phi_A({\bf r}_0)$ obtained from case A.  Then, the charges and 
potentials in case B can be summarized as

B: $\left\{ q_0,\ V_0;\ \Delta Q_i,\ V_i = 0,\ i = 1,...,n \right\}$.

We solve the electrostatics problem for a third case,

C:  $\left\{ q'_0 = 0,\ V'_0({\bf r}_0);\ Q_1,\ V'_1 = 1;\
\Delta Q_i = 0,\ V'_i = 0,\ i = 2,...,n. \right\}$,
\hfill\break\noindent
in which conductor 1 is held at unit potential, the charges on all other 
conductors
at zero, and all other conductors are grounded except for the tiny conductor at
position ${\bf r}_0$.  Again, we solve this problem as in case A, first ignoring
the tiny conductor, then evaluating $V'_0$ as $\phi_C({\bf r}_0)$.

The reciprocation theorem (\ref{eq1a}) applied to cases B and C implies that
\begin{equation}
0 = q_0 V'_0 + \Delta Q_1 \cdot 1.
\label{eq1e}
\end{equation}
The current that moves off electrode 1 in case B is therefore,
\begin{equation}
I_1 = - {d \Delta Q_1 \over dt} = q_0 {d V'_0({\bf r}_0) \over dt}
= q_0 \nabla V'_0({\bf r}_0) \cdot {d {\bf r_0} \over dt}
= - q_0 {\bf E}_{\rm w} \cdot {\bf v},
\label{eq1f}
\end{equation} 
where the velocity {\bf v} of the charge is determined using the fields from
case A, and 
\begin{equation}
{\bf E}_{\rm w} = - \nabla V'_0({\bf r}_0) = - \nabla \phi_C({\bf r}_0)
\label{eq1g}
\end{equation}
is called the weighting field.  For the case of two conductors (plus charge 
$q_0$) one of which is grounded, the weighting field is the same as the field 
from case A, but in general they are distinct.

As the present problem involves only two conductors, you may wish to find a
solution that does not appear to use the initially cumbersome machinery of the
reciprocation theorem.

\section{Solution}

The form $I(t)$ can also be found without using the reciprocation theorem, so
we illustrate that first.

\subsection{Elementary Solution for $I(t)$}

The current that flows off the anode is equal to minus the rate of change of the
charge $q(t) < 0$ that remains on the anode as the positive ions of
total charge $q_0$ move outward according to $r(t)$.  

The key to an elementary solution is that although the positive ions occupy a 
very small volume around the point
$(r,\theta,z) = (r(t),0,0)$ in cylindrical coordinates, the charge they induce 
on
the cathode is exactly the same as if those ions were uniformly spread out over
a cylinder of radius $r$.

Because the superposition principle holds in electrostatics, the problem of
the chamber with voltage $V$ on the anode plus ions at a fixed position
between the anode and cathode can be separated into two parts.  
First, an empty chamber with voltage $V$
on the anode, and second, a grounded chamber with positive ions inside.
[That is, we decompose the problem into cases A and B of sec.~1.1 even though
we won't use the reciprocation theorem here.]

For the second part, the radial electric field in the region $a < r < r(t)$ can
be calculated from the charge $q$ on the anode as
\begin{equation}
E(r) =  {2 q(t)  \over r l}\, ,
\label{eq2}
\end{equation}
using Gauss' Law, where $l \gg b$ is the length of the cylinder.
Similarly, the electric field in the
region $r(t) < r < b$ is
\begin{equation}
E(r) =  {2 (q_0 + q(t))  \over r l}\, .
\label{eq3}
\end{equation}

The potential difference between the inner and outer cylinder must be zero. 
Hence,
\begin{equation}
0 =  {2 q(t)  \over l} \int_a^{r(t)} {dr \over r} 
+ {2 (q_0 + q(t))  \over l} \int_{r(t)}^b {dr \over r} 
= {2 q_0 \over l} \ln {b \over r(t)} + {2 q(t) \over l} \ln{b \over a}\, ,
\label{eq4}
\end{equation}
and so
\begin{equation}
q(t) = - q_0 {\ln(b/r(t)) \over \ln(b/a)}.
\label{eq5}
\end{equation}
The current is
\begin{equation}
I(t) = - \dot q(t) = - {q_0 \over \ln(b/a)} {v(t) \over r(t)}.
\label{eq6}
\end{equation}

To calculate the dynamical quantities $r(t)$ and $v(t)$, 
we must return to the full problem of the
ions in a chamber with voltage $V$.  The electric field in the chamber
is only slightly perturbed by the presence of the ions, and so is given by
\begin{equation}
E(r) = {V \over r \ln(b/a)}\, .
\label{eq7}
\end{equation}
According to (\ref{eq1}), the positive ions have velocity
\begin{equation}
v(r) = {\mu V \over r \ln(b/a)}\, ,
\label{eq8}
\end{equation}
which integrates to give
\begin{equation}
r^2(t) = a^2 + {2 \mu V \over \ln(b/a)} t.
\label{eq9}
\end{equation}

Inserting (\ref{eq8}-\ref{eq9}) in (\ref{eq6}), we find
\begin{equation}
I(t) 
= - {q_0 \over 2 t_0 \ln(b/a)} {1 \over 1 + t/t_0}\, ,
\label{eq10}
\end{equation}
where
\begin{equation}
t_0 = {a^2 \ln(b/a) \over 2 \mu V}\, .
\label{eq11}
\end{equation}
The idealized current pulse has a very sharp rise, and falls off rapidly over
characteristic time $t_0$, which is about 20 nsec in typical straw tube
chambers.

\subsection{$I(t)$ via Reciprocity}

Referring to the prescription in sec.~1.1,
we first solve case C, in which the inner electrode is at unit potential and
the outer electrode is grounded.  We quickly find that
\begin{equation}
V_C(r) = { \ln (b/r) \over \ln (b/a)}\, .
\label{eq12}
\end{equation}
According to (\ref{eq1f}), the current off the inner electrode is therefore,
\begin{equation}
I(t) = - q_0 {d V_C \over dr} v(r) 
= - {q_0 \over \ln (b/a)} {v(t) \over r(t)}\, ,
\label{eq13}
\end{equation} 
as previously found in (\ref{eq6}).  We again solve for $v$ and $r(t)$ as in
(\ref{eq7}-\ref{eq9}), which corresponds to the use of case A,
 to obtain the solution (\ref{eq10}-\ref{eq11}).

\subsection{The Charge Distribution $q(z)$ on the Cathode}

The more detailed question as to the longitudinal charge distribution on the
cathode can be solved by the reciprocation method if we conceptually divide
the cathode cylinder into a ring of length $dz$ at position $z_1$ plus two
cylinders that extends to $z = \pm l/2$ where $l$ is the length of the
cylinder.  We label the ring as electrode 1 as desire the
charge $\Delta Q_1 = q(z)dz$ induced on this ring when the positive ion charge
$q_0$ is at position $(r_0,0,z_0)$ in cylindrical coordinates $(r,\theta,z)$.

According to the prescription of sec.~1.1,
\begin{equation}
\Delta Q_1 = - q_0 V_C(r_0,0,z_0),
\label{eq14}
\end{equation}
where case C now consists of a cylinder of radius $b$ grounded except for the 
ring at position $z_1$ at unit potential, and a grounded cylinder at radius 
$a$.  For
$z$ not close to the ends of the cylinder, the end surfaces $z = \pm l/2$
may be approximated as at ground potential. 

This problem is very similar to that discussed in sec.~5.36 of \cite{Smythe}.

Laplace's equation, $\nabla^2 \phi_C({\bf r}) = 0$ holds for the potential in 
the region $a < r < b$.  The problem
has azimuthal symmetry, so $\phi_C$ will be independent of $\theta$.  Since the
planes $z = \pm l/2$ are grounded, the longitudinal functions in the Fourier
series expansion,
\begin{equation}
\phi_C = \sum_n R_n(r) Z_n(z),
\label{eq15}
\end{equation}
must have the form $Z_n = \sin 2n \pi z/l$.
The equation for the radial functions $R_n(r)$ follows from Laplace's equation 
as 
\begin{equation}
{d^2 R_n \over dr^2} + {1 \over r} {d R_n \over dr} - \left( {2 n \pi \over l}
\right)^2 R_n = 0.
\label{eq16}
\end{equation}
The solutions of this are the modified Bessel functions of order zero,
$I_0(2n \pi r/l)$ and $K_0(2n \pi r/l)$.  Both of these are finite on the
interval $a < r < b$, so the expansion (\ref{eq15}) will include them both.

The boundary condition that $\phi_C(a,\theta,z) = 0$ is satisfied by the
expansion
\begin{equation}
\phi_C = \sum_n A_n { 
{I_0(2n \pi r / l) \over I_0(2n \pi a / l)}
- {K_0(2n \pi r / l) \over K_0(2n \pi a / l)} \over
{I_0(2n \pi b / l) \over I_0(2n \pi a / l)}
- {K_0(2n \pi b / l) \over K_0(2n \pi a / l)} } \sin{2 n \pi z \over l}\, ,
\label{eq17}
\end{equation}
where the form of the denominator is chosen to simplify the evaluation of the
boundary condition at $r = b$.  Here, $\phi_C = 0$, except of an interval $dz$
long about $z$ where it is unity.  Hence, the Fourier coefficients are
\begin{equation}
A_n = {2 \over l} \sin{2 n \pi z_1 \over l} dz.
\label{equation}
\end{equation}

In sum, the charge distribution q(z) on the cathode at radius $b$ due to
positive charge $q_0$ at $(r_0,0,z_0)$ follows from (\ref{eq14}) and 
(\ref{eq16}-\ref{eq17}) as
\begin{equation}
q(z) = - {2 q_0 \over l} \sum_n { 
{I_0(2n \pi r_0 / l) \over I_0(2n \pi a / l)}
- {K_0(2n \pi r_0 / l) \over K_0(2n \pi a / l)} \over
{I_0(2n \pi b / l) \over I_0(2n \pi a / l)}
- {K_0(2n \pi b / l) \over K_0(2n \pi a / l)} } 
\sin{2 n \pi z \over l} \sin{2 n \pi z_0 \over l}\, .
\label{eq18}
\end{equation}
A numerical evaluation of (\ref{eq18}) is illustrated in Fig.~1.  As is to be
expected, the induced charge distribution on the cathode has characteristic 
width of 
order $b - r_0$, the distance of the positive charge from the cathode.

\begin{figure}[htp]  
\begin{center}
\includegraphics[width=4in, angle=0, clip]{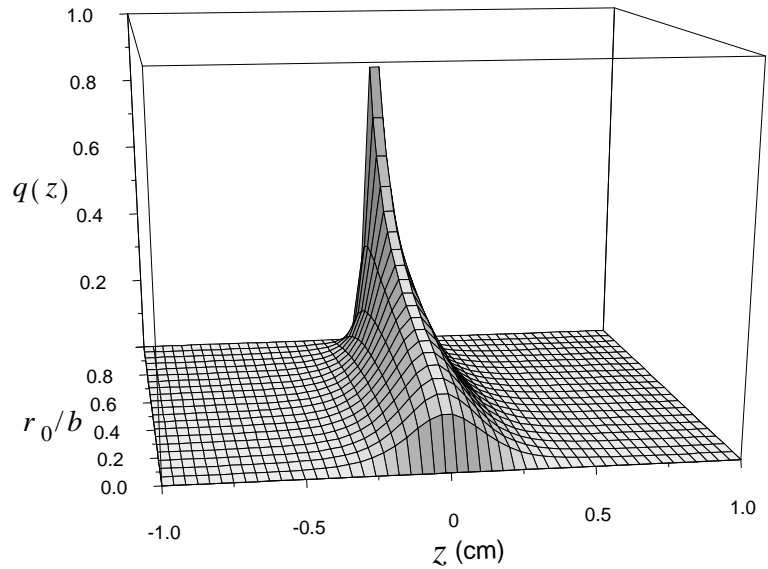}
\parbox{5.5in} 
{\caption[ Short caption for table of contents ]
{\label{fig1} The induced charge distribution (\ref{eq18}) on the cathode of a 
straw tube chamber of radius $b = 0.25$ cm due to positive ion charge at 
radius $r_0$.
}}
\end{center}
\end{figure}

\end{document}